\documentclass[aps, prb, twocolumn, a4paper, superscriptaddress]{revtex4-2}

\usepackage{graphicx} 
\usepackage{dcolumn} 
\usepackage{amsmath,bm} 
\usepackage{color} 
\usepackage{gensymb} 
\usepackage{url}

\begin{document}

\title{Megahertz dynamics in skyrmion systems probed with muon-spin relaxation}

\author{T.~J.~Hicken}
\affiliation{Centre for Materials Physics, Durham University, Durham, DH1 3LE, United Kingdom}
\author{M.~N.~Wilson}
\affiliation{Centre for Materials Physics, Durham University, Durham, DH1 3LE, United Kingdom}
\author{K.~J.~A.~Franke}
\altaffiliation{Present address: Department of Materials Science and Engineering, University of California, Berkeley, Berkeley, CA 94720, United States of America}
\altaffiliation{Other address: School of Physics and Astronomy, University of Leeds, Leeds, LS2 9JT, United Kingdom}
\affiliation{Centre for Materials Physics, Durham University, Durham, DH1 3LE, United Kingdom}
\author{B.~M.~Huddart}
\author{Z.~Hawkhead}
\affiliation{Centre for Materials Physics, Durham University, Durham, DH1 3LE, United Kingdom}
\author{M.~Gomil{\v s}ek}
\altaffiliation{Present address: Jo{\v z}ef Stefan Institute, Jamova c. 39, SI-1000 Ljubljana, Slovenia}
\affiliation{Centre for Materials Physics, Durham University, Durham, DH1 3LE, United Kingdom}
\author{S.~J.~Clark}
\affiliation{Centre for Materials Physics, Durham University, Durham, DH1 3LE, United Kingdom}
\author{F.~L.~Pratt}
\affiliation{ISIS Pulsed Neutron and Muon Facility, STFC Rutherford Appleton Laboratory, Harwell Oxford, Didcot, OX11 OQX, United Kingdom}
\author{A.~{\v S}tefan{\v c}i{\v c}}
\altaffiliation{Present address: Electrochemistry Laboratory, Paul Scherrer Institut, CH-5232 Villigen PSI, Switzerland}
\affiliation{Department of Physics, University of Warwick, Coventry, CV4 7AL, United Kingdom}
\author{A.~E.~Hall}
\affiliation{Department of Physics, University of Warwick, Coventry, CV4 7AL, United Kingdom}
\author{M.~Ciomaga~Hatnean}
\affiliation{Department of Physics, University of Warwick, Coventry, CV4 7AL, United Kingdom}
\author{G.~Balakrishnan}
\affiliation{Department of Physics, University of Warwick, Coventry, CV4 7AL, United Kingdom}
\author{T.~Lancaster}
\affiliation{Centre for Materials Physics, Durham University, Durham, DH1 3LE, United Kingdom}

\date{\today}

\begin{abstract}
  We present longitudinal-field muon-spin relaxation (LF $\mu$SR) measurements on two systems that stabilize a skyrmion lattice (SkL): Cu$_2$OSeO$_3$, and Co$_x$Zn$_y$Mn$_{20-x-y}$ for $(x,y)~=~(10,10)$, $(8,9)$ and $(8,8)$.
  We find that the SkL phase of Cu$_2$OSeO$_3$ exhibits emergent dynamic behavior at megahertz frequencies, likely due to collective excitations, allowing the SkL to be identified from the $\mu$SR response.
  From measurements following different cooling protocols and calculations of the muon stopping site, we suggest that the metastable SkL is not the majority phase throughout the bulk of this material at the fields and temperatures where it is often observed.
  The  dynamics of bulk Co$_8$Zn$_9$Mn$_3$ are well described by $\simeq~2$~GHz excitations that reduce in frequency near the critical temperature, while in Co$_8$Zn$_8$Mn$_4$ we observe similar behavior over a wide range of temperatures, implying that dynamics of this kind persist beyond the SkL phase.
\end{abstract}
\maketitle

\section{Introduction}

\begin{figure}
	\centering
	\includegraphics[width=0.9\linewidth]{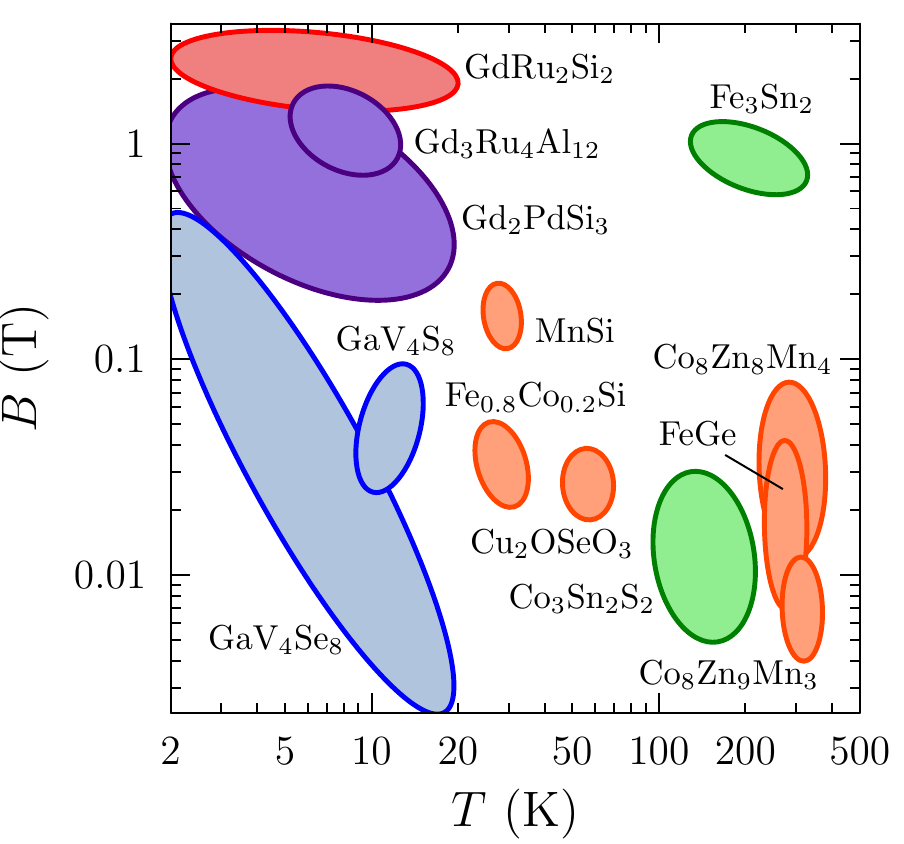}
	\caption{Skyrmion phase diagram of bulk materials. Stabilization of the skyrmion phase occurs through various mechanisms: (i) competition between the exchange and Dzyaloshinskii–Moriya interactions, leading to a Bloch (orange) or N{\'e}el (blue) skyrmion lattice, (ii) competition between exchange and uniaxial anisotropy (green), (iii) geometric frustration (purple), or (iv) interplay of RKKY and four-spin interactions (red). Phase boundaries taken from Refs.~\cite{muhlbauer2009skyrmion,munzer2010skyrmion,wilhelm2011precursor,seki2012observation,tokunaga2015new,kezsmarki2015neel,karube2016robust,hou2017observation,bordacs2017equilibrium,kurumaji2019skyrmion,hirschberger2019skyrmion,wu2020observation,khanh2020nanometric}.}
	\label{fig:skyrmiondiagram}
\end{figure}

The skyrmion has been the attention of much recent research~\cite{lancaster2019skyrmions,everschor2018perspective} due to its potential for future spintronic applications~\cite{finocchio2016magnetic,fert2017magnetic}.
Several mechanisms can lead to the stabilization of a skyrmion spin texture, with examples in thin films, multilayer stacks and bulk materials~\cite{lancaster2019skyrmions,everschor2018perspective}; the region of stability of the skyrmion phase in the $B$-$T$ phase diagram is quite different in different systems (Fig.~\ref{fig:skyrmiondiagram})~\cite{muhlbauer2009skyrmion,munzer2010skyrmion,wilhelm2011precursor,seki2012observation,tokunaga2015new,kezsmarki2015neel,karube2016robust,hou2017observation,bordacs2017equilibrium,kurumaji2019skyrmion,hirschberger2019skyrmion,wu2020observation,khanh2020nanometric}.
For applications it is important to understand the spin dynamics of skyrmions, which are most commonly studied in systems which host a skyrmion lattice (SkL).
Originally detected using microwave spectroscopy techniques in Cu$_2$OSeO$_3$~\cite{onose2012observation}, three modes in the GHz regime are identified as excitations of the Bloch skyrmion: counterclockwise, breathing, and clockwise modes~\cite{schwarze2015universal}. 
In addition, some SkL-hosting materials show other collective excitations, e.g. THz excitations in Cu$_2$OSeO$_3$ due to spin excitations in high-energy magnon bands~\cite{gnezdilov2010magnetoelectricity,miller2010magnetodielectric,ozerov2014establishing,romhanyi2014entangled}.

In general, ordered magnets host diffusive and propagating magnetic excitations; we therefore expect excitations over a wide range of frequencies.
Despite this, there are few reports on the excitation spectra of SkL-hosting materials in the MHz regime.
One technique that can probe this regime, which is sensitive to dynamics of the internal magnetic field, is longitudinal-field muon-spin relaxation (LF $\mu$SR).
LF $\mu$SR is a technique with a unique time-window, sensitive to dynamics between approximately 10~kHz and 1~THz~\cite{yaouanc2011muon}.
It has previously been applied to exponentially correlated fluctuations of the dense array of moments found in typical magnetic materials~\cite{yaouanc2011muon} and to more complex dynamic behaviour such as diffusive and ballistic transport in spin chains~\cite{pratt2006low,huddart2020magnetic}, correlated fluctuations in metallic ferromagnets~\cite{hayano1978observation}, and soliton motion in polymers~\cite{risch1992direct}.
However, only a handful of results have been reported where LF $\mu$SR is used to study the skyrmion lattice (SkL).
Studied systems include Cu$_{2-x}$Zn$_x$OSeO$_3$~\cite{stefancic2018origin} (which hosts Bloch skyrmions) and GaV$_4$S$_{8-y}$Se$_y$~\cite{franke2018magnetic,hicken2020magnetism} (which hosts N{\' e}el skyrmions).
Similar behavior is observed in these materials, with an enhanced and broadened peak in the muon-spin relaxation rate found at temperatures just below the critical temperature $T_\text{c}$ at those external magnetic fields that stabilize the SkL.

Here we investigate the LF $\mu$SR response in two materials that host a Bloch SkL.
We study two different systems with contrasting crystal symmetries in which the non-centrosymmetric crystal structure leads to a bulk Dzyaloshinskii–Moriya interaction, thus stabilizing the SkL through competition with symmetric exchange.
The first is Cu$_2$OSeO$_3$~\cite{seki2012observation}, an insulating, multiferroic ferrimagnet which crystallizes in the P2$_1$3 structure.
The stability and extent of the skyrmion lattice in Cu$_2$OSeO$_3$ can be controlled both with an externally applied electric field ($E$-field)~\cite{white2012electric} and through chemical substitution of the magnetic Cu ions~\cite{birch2019increased}.
The second is Co$_x$Zn$_y$Mn$_{20-x-y}$~\cite{tokunaga2015new}, which is a metallic system with the $\beta$-Mn structure, known for its chemical substitutional site disorder.
The magnetic properties of the system change significantly with $x$ and $y$ and the series is of particular interest in those compositions that host a SkL above room temperature.
In both systems we are able to use muons to observe a dynamic response on the MHz timescale that is unique to fields which stabilize the SkL.

The paper is structured as follows: in Sec.~II we describe the experimental and analytical procedures used; in Sec.~III we first probe Cu$_2$OSeO$_3$ with $\mu$SR and analyze the results with support from density functional theory (DFT) calculations of the muon stopping site, before turning to Co$_x$Zn$_y$Mn$_{20-x-y}$, where we present muon stopping site calculations and investigate three different compositions.
Additional details can be found in the Supplemental Material~\cite{si}.

\section{Experimental}

Cu$_2$OSeO$_3$ samples were synthesized as detailed in Ref.~\cite{stefancic2018origin}, and polycrystalline Co$_x$Zn$_y$Mn$_{20-x-y}$ boules were synthesized as detailed in the Supplemental Material~\cite{si}.
In a LF $\mu$SR experiment spin-polarized positive muons are implanted in a sample in the presence of an external magnetic field parallel to the initial muon-spin direction~\cite{blundell1999spin,si}.
Implanted muons interact with the local magnetic field at the muon site, which is a sum of the external and internal fields.
By measuring the decay of the polarization of the spin of the muon ensemble, one reveals information about both the static and dynamic properties of the local magnetism at the muon site.
In the fast-fluctuation regime, typical for an ordered magnet, this relaxation is exponential with a relaxation rate
\begin{equation}\label{eqn:redfield}
	\lambda~=~\frac{2\Delta^2\nu}{\omega_0^2+\nu^2} ,
\end{equation}
where $\nu$ is the characteristic frequency of the field fluctuations, $\Delta~=~\gamma_\mu\sqrt{\langle B^2\rangle}$ is the amplitude of the field fluctuations, and $\omega_0~=~\gamma_\mu B_\text{ext}$ is the precession frequency of a muon with gyromagnetic ratio $\gamma_\mu~=~2\pi\times$135.5~MHz~T$^{-1}$ in the external field $B_\text{ext}$.
The fluctuations themselves can be described by a spectral density $J(\omega)$, which represents the Fourier transform of the autocorrelation function of the magnetic field at the muon site(s).
In cases where $J(\omega)$ is broad in frequency, the muon spin polarization will be most effectively relaxed by the part of the spectral density at frequencies close to $\omega_{0}~=~\gamma_\mu B_\text{ext}$, which typically lies in the MHz regime for values of $B_\text{ext}$ applied in our measurements.

\section{Results \& Discussion}

\subsection{Cu$_{\bm{2}}$OSeO$_{\bm{3}}$}

\begin{figure}
	\centering
	\includegraphics[width=\linewidth]{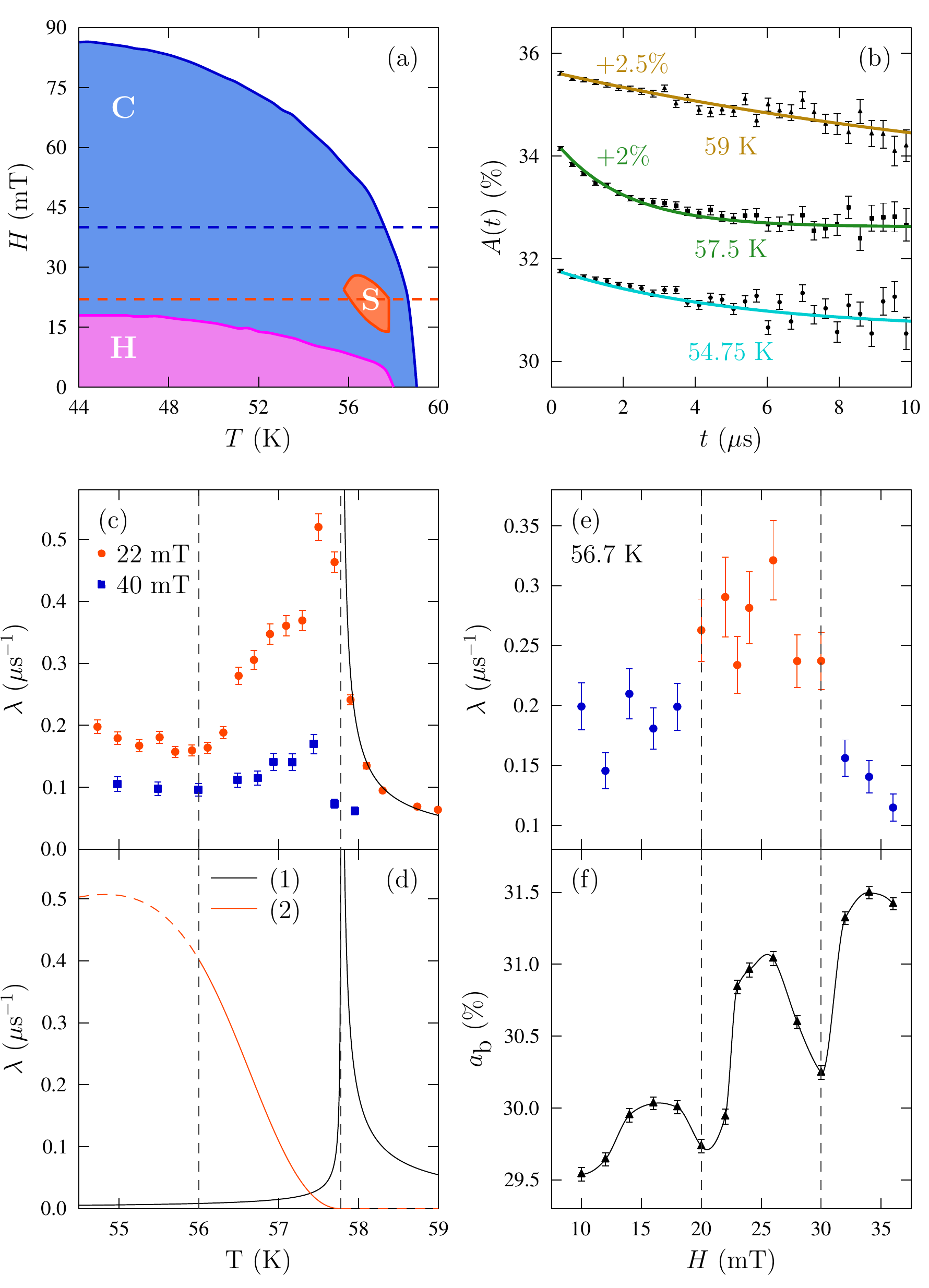}
	\caption{(a) Phase diagram of Cu$_2$OSeO$_3$, showing conical (C), helical (H) and skyrmion (S) phases, reproduced from Ref.~\cite{stefancic2018origin}. Orange color: fields at which the SkL is stabilized, blue: fields which only stabilize C order below $T_\text{c}$, and pink: H order. (b) Example LF $\mu$SR spectra for Cu$_2$OSeO$_3$ measured in  $B=22$~mT. For clarity, some data are shown with vertical offsets. (c) Extracted values of $\lambda$. (d) Simulations of contributions to $\lambda$ due to: (1) critical slowing down of magnetic fluctuations near $T_\text{c}$; (2) reduction in frequency of GHz spectral density. Orange dashed line indicates the value of $\lambda$ one would obtain if the SkL was stabilized at those temperatures. Vertical dashed lines indicate the location of the SkL at 22~mT from AC susceptibility. (e) Extracted values of $\lambda$ from a field scan at $T=56.7$~K, with (f) accompanying baseline amplitude (solid line is a guide to the eye). Dashed lines indicate the location of the SkL at 56.7~K.}
	\label{fig:cu2oseo3_sc}
\end{figure}

LF $\mu$SR measurements on a mosaic of single-crystals of Cu$_2$OSeO$_3$ were performed on warming, after cooling in zero applied magnetic field (ZFC).
Temperature scans were performed at $B_\text{ext}~=~22$~mT, which stabilizes the SkL state between $\approx~56$~K and $\approx~58$~K, and at $B~=~40$~mT, which does not stabilize a SkL at any temperature [Fig.~\ref{fig:cu2oseo3_sc}(a)].
Example spectra are shown in Fig.~\ref{fig:cu2oseo3_sc}(b), where the asymmetry decays monotonically, with an exponential decay typical of relaxation due to the dynamics of a dense array of fluctuating local moments.
The spectra are well described at all measured temperatures using a relaxation function
\begin{equation}\label{eqn:1lor}
A\left(t\right)~=~a_\text{r}e^{-\lambda t} + a_\text{b} ,
\end{equation}
where the component with amplitude $a_\text{r}$ captures the contribution from muons stopping in the sample with their spin initially aligned along the local magnetic field, and the baseline amplitude $a_\text{b}$ accounts for muons that stop outside of the sample or at positions in the material where a fluctuating field does not dephase them. 
The relaxing amplitude $a_\text{r}$ increases as the temperature is raised through the ordering temperature $T_\text{c}$.
To model this, we constrain $a_\text{r}$ to 
\begin{equation}\label{eqn:sigmoid}
a_\text{r}~=~a_\text{r}^0 + L \left[1+e^{-k\left(T-T_\text{c}\right)}\right]^{-1} , 
\end{equation}
which allows us to extract $T_\text{c}$ independently of $\lambda$.
(Extracted values of $T_\text{c}$ agree well with those from AC susceptibility.)
There is no temperature dependence to $a_\text{b}$, leaving only $\lambda$ [Fig.~\ref{fig:cu2oseo3_sc}(c)] varying in our fits.

There are striking differences in the behavior of $\lambda$ between the two temperature scans.
On scanning through the fields in the $B$-$T$ phase diagram where the SkL is realized, $\lambda$ is significantly enhanced at those temperatures where the SkL phase is found, resulting in a broad shoulder above 56.5~K that terminates in a peak on the high temperature side.
No such enhancement is observed at higher fields.
This indicates significantly enhanced $J(\omega)$ around $\omega_0~\simeq~3$~MHz in the SkL phase and is similar to the behavior previously reported for both the Bloch and N{\'e}el-type SkL~\cite{stefancic2018origin,hicken2020magnetism}.
Therefore, we conclude that LF $\mu$SR has a characteristic response to the SkL, specifically an the increase in relaxation rate compared to the surrounding magnetic phases.

As variation in the amplitude of the fluctuating field $\Delta/\gamma_{\mu}$ is likely to follow the magnetization, the variation in $\lambda$ likely results from the temperature dependence of $\nu$, and could reflect:
(1) critical slowing down of the magnetic fluctuations near $T_\text{c}$, typical of a second-order phase transition; (2) reduction in frequency of the  skyrmion excitation modes near $T_\text{c}$; (3) other collective dynamics of the system occurring on the MHz timescale.

(1) Above $T_\text{c}$, the relaxation rate $\lambda$ is well described by power-law behavior~\cite{pratt2007chiral} typical of critical fluctuations in  a 3D Heisenberg magnet~\cite{pelissetto2002critical,pospelov2019non, vzivkovic2014critical,wilson2020stability} with a fluctuation time $1/\nu \propto \left|T-T_\text{c}\right|^{-w'}$ with $w'~=~0.709$, typical for a 3D Heisenberg magnet.
Below $T_\text{c}$ the same critical parameters do not account for $\lambda$ which should show a sharp rise very close to $T_\text{c}$ [Fig.~\ref{fig:cu2oseo3_sc}(d)].

(2) The skyrmion  rotational and breathing modes are expected to broaden and decrease in frequency (or soften) as $T$ approaches $T_\text{c}$ from below (see, for example, Ref.~\cite{birch2019increased}), contributing to the spectral weight  $J\left(\omega\right)$ centered around $\omega = \gamma_{\mu}B_\text{ext}$ ($2\pi\times~3$~MHz at our value $B_\text{ext}$).
Assuming that the time-dependent magnetization that results from skyrmion modes determines the relaxation, we can use typical exponents for a 3D Heisenberg model to predict
\begin{equation}\label{eqn:mode_soften}	\lambda~=~\frac{2\Delta_0^2\nu_0\left[1-(T/T_\text{c})^{3/2}\right]^{0.73}\left(1-T/T_\text{c}\right)^{1.43}}{\gamma_\mu^2B_\text{ext}^2+\nu_0^2\left(1-T/T_\text{c}\right)^{2.86}} .
\end{equation}
Cu$_2$OSeO$_3$ exhibits its lowest frequency  skyrmion mode (counterclockwise rotational) at $\nu_0~=~2.3$~GHz~\cite{garst2017collective}, giving the behavior shown in Fig.~\ref{fig:cu2oseo3_sc}(d), which  does not describe the measured data.
Note that $\nu_0~=~10$--20~GHz would be a better match to the data, but this is at least a factor of 3-4 higher than the three lowest energy modes of the SkL in Cu$_2$OSeO$_3$~\cite{garst2017collective}, but too low in frequency to be the THz excitations previously observed.

(3) Alternatively, $\lambda$ could reflect the occurrence of other low-energy, collective excitations emerging from the SkL involving individual skyrmions or from motion of the SkL (e.g.\ diffusive excitations resulting when the SkL undergoes collective motion, or where individual skyrmions are created or destroyed.)
This is plausible given that diffusive dynamics for single skyrmions~\cite{miltat2018brownian,weissenhofer2020diffusion} typically occur in the GHz regime, while the motion of Bloch points along skyrmion tubes are likely to occur at MHz frequencies.
It could also be that $\Delta$ increases near the transition owing to rapid changes in width of the local field distribution at the muon sites. In each of these cases a change in the distribution of skyrmions in the SkL is required.

To further investigate the response to the SkL, LF $\mu$SR measurements were also made as a function of increasing applied magnetic field $B$ at fixed temperature $T=56.7$~K after ZFC.
These data are also well described by Eq.~\ref{eqn:1lor}, with a field-independent amplitude $a_\text{r}$ and a baseline $a_{\mathrm{b}}$ that increases with $B$, as is often observed in LF $\mu$SR.
We again observe enhanced values of  $\lambda$ in the SkL phase, along with discontinuous behavior in $a_\text{b}$ marking the transitions in and out of the SkL phase [Fig.~\ref{fig:cu2oseo3_sc}(e--f)],  providing another method of identifying the SkL phase boundaries. 
A likely explanation of the observed behavior comes from demagnetization effects which are known to cause the magnetic transition in and out of the SkL state to occur at slightly different fields for different parts of the sample~\cite{reimann2018neutron}.
This leads to increased disorder in the field distribution at the muon sites, resulting in fewer muons stopping with their spin parallel to the local field, and hence dephasing too rapidly to be observed resulting in a loss of 
the total, and hence baseline, asymmetry.

\begin{figure}
	\centering
	\includegraphics[width=0.7\linewidth]{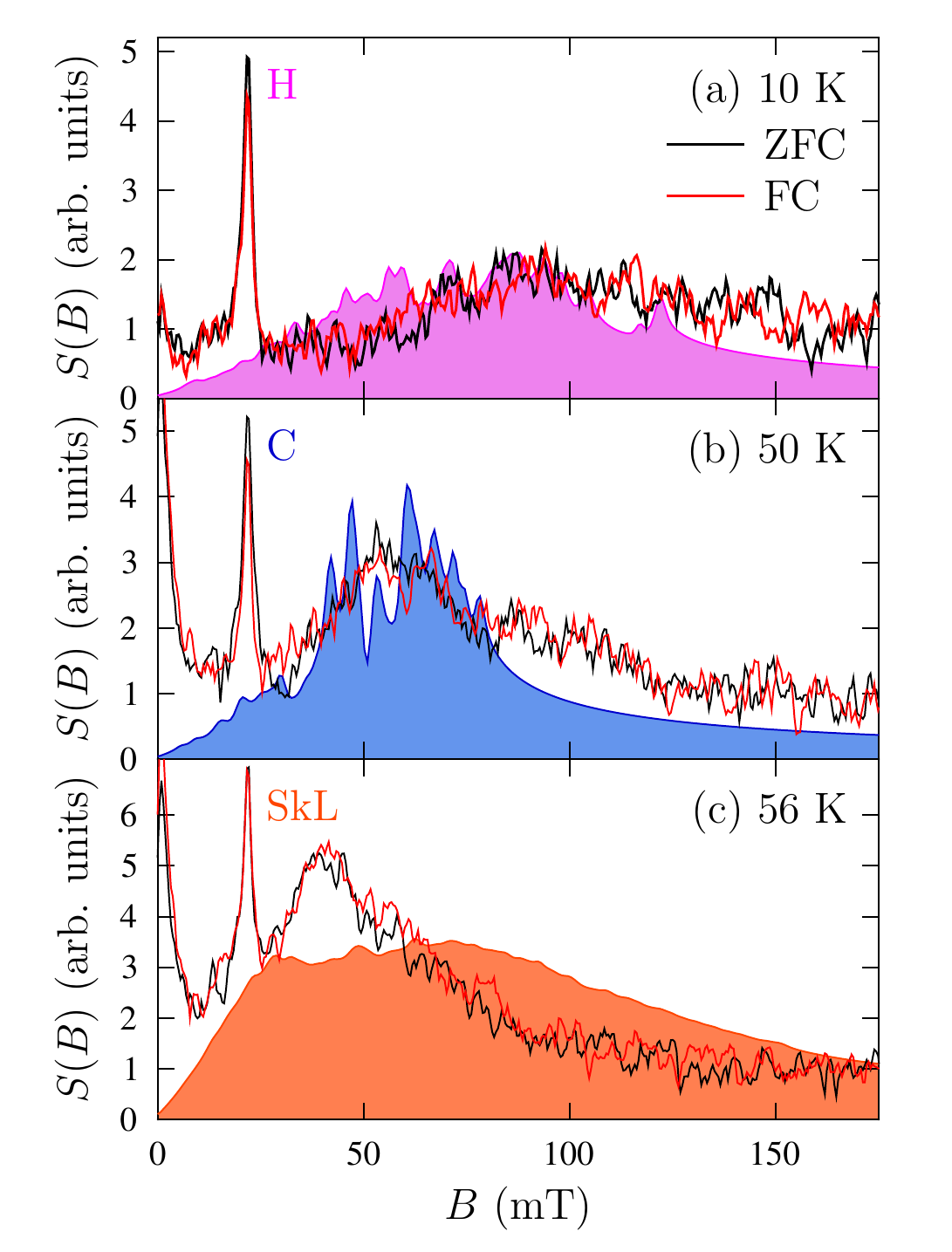}
	\caption{Internal magnetic field distributions [$S(B)$] of Cu$_2$OSeO$_3$ at various $T$ measured by TF $\mu$SR for $B_\text{ext}= 22$~mT, with comparison to simulations. Black lines: measurements performed after ZFC; red lines: after rapid FC ($\approx~17$~K/minute); solid color: simulated distributions.}
	\label{fig:cu2oseo3_ooe}
\end{figure}

We now discuss the internal field distribution, muon sites, and the possibility of observing metastable skyrmions in Cu$_{2}$OSeO$_{3}$. 
We performed TF $\mu$SR measurements on a single crystal of Cu$_2$OSeO$_3$ after ZFC and rapid cooling in applied field (FC) ($\approx~17$~K/minute).
This rapid FC is expected to stabilize metastable skyrmions at temperatures that host a conical phase for ZFC~\cite{birch2019increased}.
TF measurements are sensitive to the static internal magnetic field distribution of the sample at the muon sites, and has been shown to be sensitive to changes in the magnetic state in this material~\cite{lancaster2015transverse}.
Internal field distributions derived from TF measurements measured after both ZFC and rapid FC are compared in Fig.~\ref{fig:cu2oseo3_ooe} in an applied field of 22~mT.
There is a characteristic change in distributions for the different magnetic states~\cite{lancaster2015transverse}, which are observed after both field protocols, but no significant difference is observed between the two protocols,
suggesting that the local field distribution is similar in both cases.
(The peak observed at 22~mT at all temperatures occurs from muons stopping outside of the sample and precessing in the applied field.)

To model magnetic field distributions for the ordered states in Cu$_{2}$OSeO$_{3}$,  muon stopping sites were determined using DFT methods to relax the structure with an implanted muon~\cite{moller2013playing}.
We find different sites to those identified in Ref.~\cite{maisuradze2011mu}, which were determined by finding the minima of the electrostatic potential in the crystal with no muon present.
Our three distinct sites (of which only the two lowest energy are found to be occupied) are shown in Tab.~\ref{tab:cu2oseo3_sites}, with simulated field distributions at the muon sites shown in Fig.~\ref{fig:cu2oseo3_ooe}
(for further details see~\cite{si}).
The simulations describe the experimental results reasonably well, with the worst match being for the SkL where, on increasing field, the rapid increase followed by slow decrease of weight is captured.
but the absolute values do not agree closely.
The good match between simulation and experiment in the helical and conical state show that the static magnetism in these phases is sufficient to describe the response on the muon timescale.
The greater discrepancy between simulation and experiment in the SkL state provides further evidence, independently from the LF $\mu$SR, that there is a significant dynamic effect on the muon timescale which affects the internal magnetic field distribution of the sample.
The lack of difference in the distributions observed after ZFC and FC protocols indicates that the metastable SkL is not the majority phase through the entire sample~\cite{si}.
This could suggest that the metastable SkL is more likely to exist in particular parts of a sample, such as sample edges or defects.
A propensity to form near surfaces would also explain the lack of any muon signal, since the muons penetrate several microns into the sample in the measurements.

\begin{table}[b]
	\begin{tabular}{c|c|c}
		Muon site & Fractional coordinates & Energy (eV) \\
		\hline
		\hline
		1 & (0.906, 0.590, 0.100) & 0.00 \\
		2 & (0.172, 0.365, 0.319) & 0.09 \\
		3 & (0.224, 0.670, 0.289) & 0.15 \\
	\end{tabular}
	\caption{Muon stopping sites in Cu$_{2}$OSeO$_{3}$. Energies are given relative to the lowest energy site.}
	\label{tab:cu2oseo3_sites}
\end{table}

Another potential way to manipulate the SkL density in Cu$_2$OSeO$_3$ is through the application of an $E$-field whose orientation and magnitude causes the SkL phase to exist over different $T$-ranges~\cite{white2012electric}.
We performed LF $\mu$SR on a polycrystalline pellet of Cu$_2$OSeO$_3$ with an $E$-field applied parallel/antiparallel to the externally applied magnetic field $B_\text{ext}$~\cite{si}.
Similarly, we find that in those cases where the SkL is not the majority phase its dynamic signature is not resolved, making it likely that $\mu$SR is sensitive to the SkL in this system only when it is the majority volume phase.

\subsection{Co$_{\bm{x}}$Zn$_{\bm{y}}$Mn$_{\bm{20-x-y}}$}

We now turn to the Co$_x$Zn$_y$Mn$_{20-x-y}$ system with $(x,y)~=~(10,10)$,
$(8,9)$ and $(8,8)$.
Some members of the Co$_x$Zn$_y$Mn$_{20-x-y}$ series host a SkL at or above room temperature, making the series potentially favorable for future applications.
However, crystallographic site disorder inherent in these materials presents challenges, such as the broadening, both in temperature and applied field, of the magnetic transitions due to locally different crystallographic environments throughout the sample, and dramatic effects on $T_\text{c}$ with relatively subtle changes in composition~\cite{bocarsly2019deciphering}.
By studying these three materials we can consider the effect of increasing site disorder (which mainly occurs on the 12d Wyckoff site) on the magnetism.
The level of disorder increases with decreasing $y$ until, once $y~\lesssim~7$, a spin glass ground state is realized~\cite{karube2018disordered}.
Here we study the regime where the system remains magnetically ordered.

The muon stopping sites in Co$_{10}$Zn$_{10}$, Co$_8$Zn$_9$Mn$_3$, and Co$_8$Zn$_8$Mn$_4$ were calculated using DFT~\cite{si}, with results presented in Tab.~\ref{tab:coznmn_sites}.
The environment of the muon affects its energy, so due to the significant site disorder, each site has a range of energies (depending on the atoms near the site for the particular simulated structure).
This will affect which sites are occupied.

\begin{table}[b]
	\begin{tabular}{c|c|c}
		Muon site & Fractional coordinates & Typical energy range (eV) \\
		\hline
		\hline
		1 & (0.179, 0.571, 0.319) & 0.00--0.98 \\
		2 & (0.344, 0.398, 0.337) & 0.31--1.03 \\
		3 & (0.426, 0.568, 0.073) & 0.42--0.97 \\
	\end{tabular}
	\caption{Muon stopping sites in Co$_x$Zn$_y$Mn$_{20-x-y}$. Typical energies of the site are given relative to the lowest energy site. The ranges reflect the fact that the local environment of each site affects the energy. For sites 2 and 3 the energy ranges only apply for 20\% Mn concentration and below, above this concentration the sites are not realized in the calculations.}
	\label{tab:coznmn_sites}
\end{table}

\begin{figure}
	\centering
	\includegraphics[width=\linewidth]{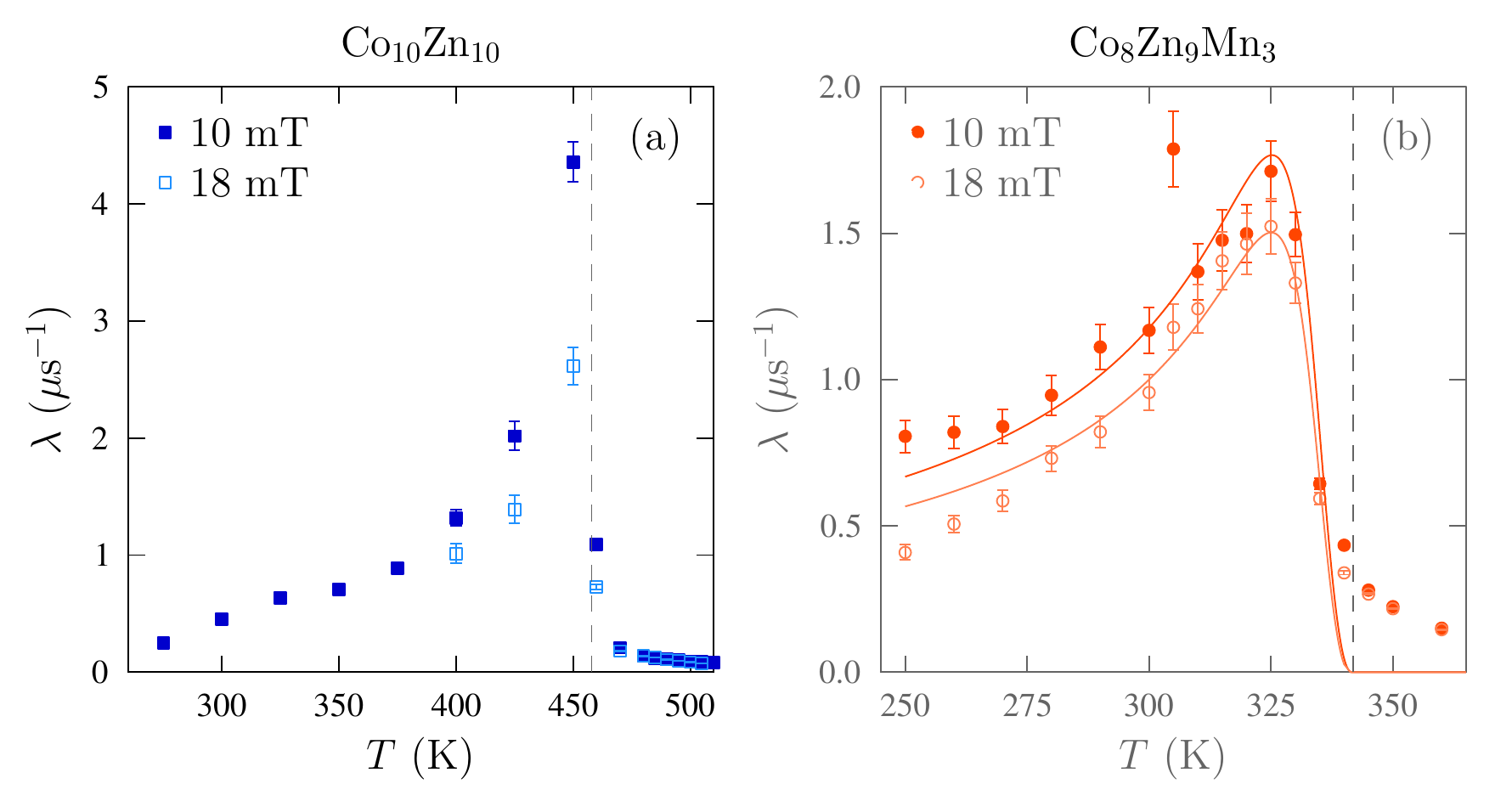}
	\caption{Extracted values of $\lambda$ from fitting LF $\mu$SR measurements of (a) Co$_{10}$Zn$_{10}$and (b) Co$_8$Zn$_9$Mn$_3$. The dashed lines indicate the average value of $T_\text{c}$ in each sample according to the relaxing amplitude. Fits in (b) are described in the text.}
	\label{fig:coznmn}
\end{figure}

We first consider the LF $\mu$SR response of the parent compound, Co$_{10}$Zn$_{10}$, which has not been reported to stabilize a SkL.
Exponential decay of the asymmetry is seen at all measured temperatures and magnetic fields.
A weak, temperature-independent relaxation is observed on the baseline with a rate consistent with Ag ($\lambda_\text{b}~=~0.0026(2)$~$\mu$s$^{-1}$).
The data are fitted to the function
\begin{equation}
A\left(t\right)~=~a_\text{r}e^{-\lambda t} + a_\text{b}e^{-\lambda_\text{b} t}.
\end{equation}
The relaxing amplitude is again constrained to follow Eqn.~\ref{eqn:sigmoid} and the resulting relaxation rate $\lambda$ is shown in Fig.~\ref{fig:coznmn}(a).
We find that measurements at two longitudinal fields have similar temperature dependence, with the overall shape of $\lambda$ is reminiscent of that measured for Cu$_2$OSeO$_3$ at 40~mT [Fig.~\ref{fig:cu2oseo3_sc}(c)], i.e.\ outside the skyrmion phase, where there is also a transition from the conical to paramagnetic phase. 
There is no evidence for any additional dynamics at either field, with the sharp peak likely occurring due to critical slowing down of the magnetic fluctuations as the phase transition from the conical to paramagnetic phase is approached [c.f. Fig.~\ref{fig:cu2oseo3_sc}(d)].

Next we discuss Co$_8$Zn$_9$Mn$_3$, a composition which can stabilize not only a SkL, but also a meron-antimeron spin texture.
Both of these textures have been observed in thin plates~\cite{yu2018transformation}, with evidence for the SkL in the bulk consisting of magnetization and magnetic entropy measurements~\cite{bocarsly2019deciphering}.
We made measurements in two applied fields: $B_\text{ext}~=~10$~mT, which stabilizes a SkL just below $T_\text{c}$ in bulk samples $(321~\lesssim~T~\lesssim~326$~K), and 18~mT, which gives a field-polarized magnetic state.
The spectra decay exponentially at all temperatures and fields and are fitted using the same method as above, with $\lambda$ shown in  Fig.~\ref{fig:coznmn}(b).
The temperature dependence of $\lambda$ is different to that found in Co$_{10}$Zn$_{10}$: the peak for Co$_8$Zn$_9$Mn$_3$  is significantly broadened with the peak in $\lambda$ occurring significantly below the obtained $T_\text{c}$, with similar behavior seen at both a field that is expected to stabilize the SkL and one that is not.
The relaxation rate $\lambda$ at both fields is well described by Eqn.~\ref{eqn:mode_soften} involving coupling to GHz excitations; although we assume 3D Heisenberg scaling parameters, the result is robust with a different choice of parameters.
Further, critical behavior of both $\Delta$ and $\nu$ are required to well describe the data; critical behavior of one parameter alone cannot describe these data.
The fits shown in Fig.~\ref{fig:coznmn}(b) suggest $\Delta_{T=0}~\simeq$~0.01--0.02~mT, and $\nu_{T=0}~\simeq~2$~GHz, so that this frequency can be identified with the characteristic excitations in this regime.
The fitted frequency is very similar to those found for other SkL systems~\cite{garst2017collective} and suggest that there are dynamics occurring over a range of fields with spectral weight that decreases in frequency with increasing temperature, passing through the frequency range that $\mu$SR is sensitive to just below $T_\text{c}$.
The fits to the model are best above $T~\simeq~280$~K, which is far greater in extent than the reported stability region of the SkL in bulk samples, but where the SkL and meron-antimeron states are reported in thin plates~\cite{yu2018transformation}.
The wide range of fields over which we detect enhanced dynamics in these bulk samples and the contrast in the extent of the SkL in plates might therefore suggest that the decisive mechanism determining the extent of the phase diagram in the thin plate samples of Co$_8$Zn$_9$Mn$_3$ is confinement.

\begin{figure}
	\centering
	\includegraphics[width=\linewidth]{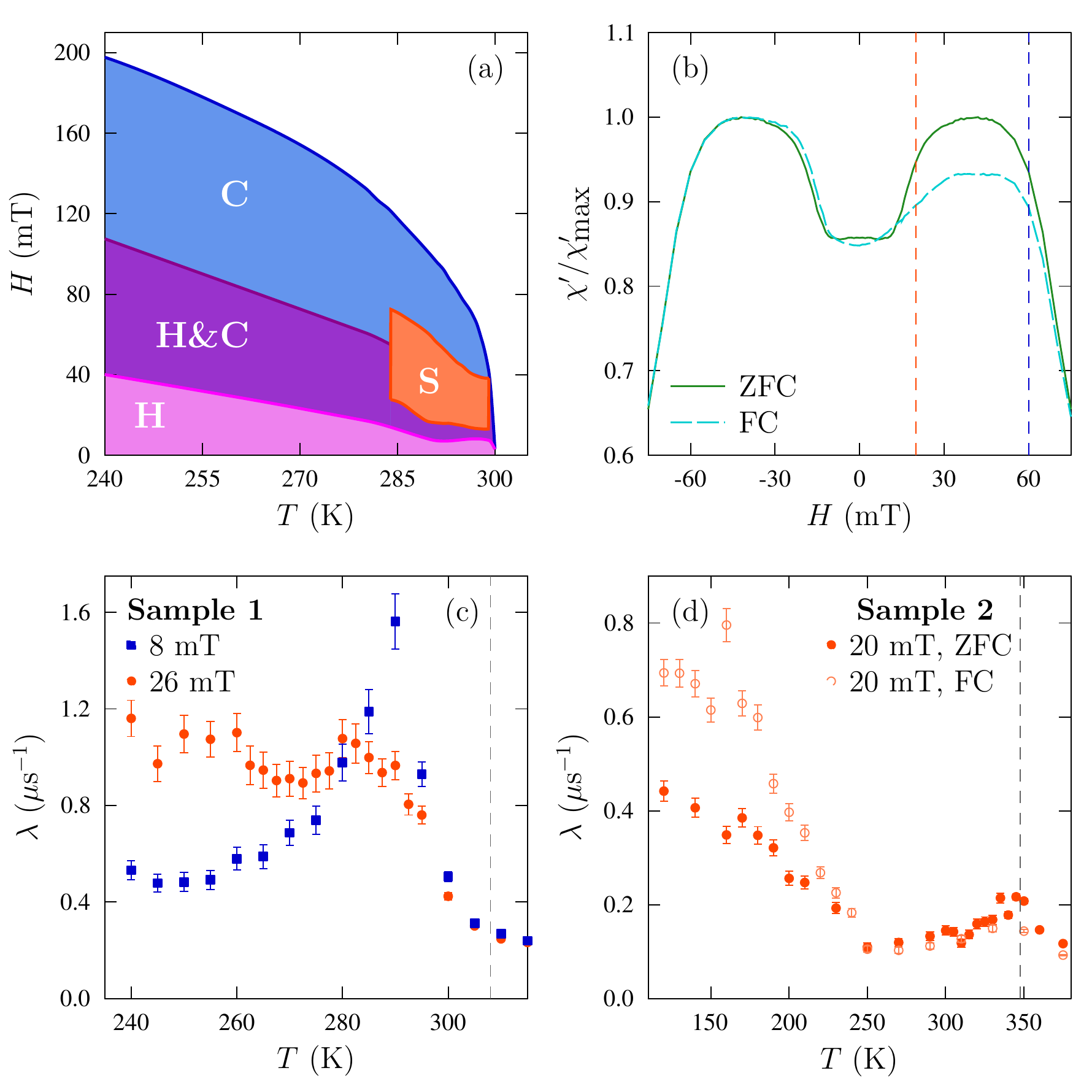}
	\caption{(a) Representative phase diagram of Co$_8$Zn$_8$Mn$_4$, showing helical (H), conical (C) and skyrmion lattice (S) phases, as well as a region of coexistence \cite{karube2016robust}. (b) AC susceptibility measurements at 250~K after ZFC and FC (in 15~mT) on one sample of Co$_8$Zn$_8$Mn$_4$, indicating a metastable SkL after FC. Fields measured with $\mu$SR in (d) and Fig.~S5 are indicated with dashed lines. (c) and (d) Relaxation rate $\lambda$ from LF $\mu$SR measurements on two different samples, with $T_\text{c}$ indicated. In (d) different field protocols are employed.}
	\label{fig:co8zn8mn4}
\end{figure}

We now turn to Co$_8$Zn$_8$Mn$_4$, which hosts a SkL around room temperature, as shown in Fig.~\ref{fig:co8zn8mn4}(a)~\cite{karube2016robust}, with the exact location of the SkL phase dependent on the precise level of Mn present.
LF $\mu$SR measurements on sample 1 again show exponential relaxation. 
The same fitting procedure is employed as above, with $\lambda$ shown in Fig.~\ref{fig:co8zn8mn4}(e).
At $B_\text{ext}~=~8$~mT, which is not expected to cut through the SkL, the peak in $\lambda$ looks typical of previous measurements that do not cut through the SkL, although the peak is well below $T_\text{c}$.
The behavior at 26~mT (which does cut through the SkL) is more unusual, with a flattened, broad maximum, and enhanced values of  $\lambda$ observed over a range of temperatures.
The suppressed peak at $T_\text{c}$ is consistent with different grains of the sample undergoing a transition at slightly different temperatures, caused by slightly varied compositions across parts of the sample.
(Mn metal has considerable vapor pressure at 1025~\degree C, meaning that Mn can migrate toward the surface of the melt during sample synthesis, forming a gradient in composition as observed in Ni$_2$MnGa~\cite{schlagel2000chemical}.)

We find that the enhanced relaxation rate in Co$_8$Zn$_8$Mn$_4$ is found at those fields that stabilize the SkL in the $B$-$T$ phase diagram, even at temperatures lower than those that stabilize the SkL state.
It is notable both that similar behavior is seen in lightly-substituted GaV$_4$S$_{8-y}$Se$_y$~\cite{hicken2020magnetism}, and that individual skyrmion formation has been reported in MnSi above $T_{\mathrm{c}}$ at those fields that stabilize the SkL~\cite{kindervater2019weak}.
These three systems, all crystallographically distinct (and, in the case of GaV$_4$S$_{8-y}$Se$_y$, hosting a different type of skyrmion), all therefore suggest the important parameter for skyrmion creation in the Hamiltonian is the applied field, with, as is the current consensus~\cite{everschor2018perspective,lancaster2019skyrmions}, thermal fluctuations stabilizing the SkL phase.

Finally we consider the effect of a FC protocol in Co$_8$Zn$_8$Mn$_4$, which is expected to stabilize a metastable SkL over a wide range of temperature.
For these measurements we used a different polycrystalline boule (sample 2).
To confirm the existence of the metastable SkL, AC susceptibility measurements were performed and are presented in Fig.~\ref{fig:co8zn8mn4}(b); the suppression of $\chi'$, typical of the SkL, is seen over a wide range of fields when employing a FC protocol.
LF $\mu$SR measurements were performed after both ZFC and FC protocols, with $\lambda$ shown in Fig.~\ref{fig:co8zn8mn4}(d).
The  higher value of $T_\text{c}$ is likely obtained due to a subtly different composition of sample 2 compared to sample 1 used for these measurements (specifically, we expect that sample 2 is Mn deficient, with
the differences between samples likely occurring due to the different dwell times during synthesis as previously discussed.)

The data measured at $B_\text{ext}~=~60$~mT (see SI~\cite{si}) show a  peak in $\lambda$, typical of those scans that do not cut through the SkL. 
At 20~mT, where a SkL is formed just below $T_\text{c}$, similar behavior is seen as was found in Fig.~\ref{fig:co8zn8mn4}(c), with a suppressed, flattened peak at $T_\text{c}$ for both field protocols.
There is, however, an enhanced response in $\lambda$ at low $T$ for FC compared to ZFC, suggesting that the stabilization of a metastable SkL is affecting the dynamics we observe. 
Although this  contrasts with the results seen for Cu$_2$OSeO$_3$, it is consistent with the expected higher stability, and hence increased volume, of the metastable SkL in Co$_8$Zn$_8$Mn$_4$~\cite{karube2016robust,birch2019increased}.
This matches the picture where site disorder allows dynamics similar to those observed in the SkL to persist to lower temperatures, as is the case in GaV$_4$S$_{8-y}$Se$_y$~\cite{hicken2020magnetism}.
In this case, stabilization of the metastable SkL likely make the dynamics more prominent, leading to the enhanced $\lambda$ observed. 

\section{Conclusion}

In Cu$_2$OSeO$_3$ high statistics LF $\mu$SR measurements reveal complex behavior in the SkL phase, possibly reflecting diffusive excitations of the skyrmion state, either through collective motion or the creation or annihilation of skyrmions.
Through TF $\mu$SR measurements and calculation of the muon stopping sites in Cu$_2$OSeO$_3$ we suggest that the metastable SkL is unlikely to be found throughout the entire sample, and we suggest that it may be more stable at boundaries and surfaces.

A range of behavior is observed in Co$_x$Zn$_y$Mn$_{20-x-y}$.
We have shown that there are MHz dynamics in Co$_8$Zn$_9$Mn$_3$, regardless of whether the field stabilizes a SkL, that can be well described by a model involving coupling to $\simeq~2$~GHz excitations whose frequency drops near $T_\text{c}$.
In Co$_8$Zn$_8$Mn$_4$ we have shown evidence for enhanced dynamics over a wide range of temperatures when the external field is one that stabilizes the SkL.
Stabilization of a metastable SkL enhances these dynamics, likely due to a greater proportion of muons being sensitive to SkL effects.
Identifying the precise source of these MHz dynamics in SkL systems, now observed in multiple materials, should be an avenue for future research.

\section*{Acknowledgments}
Part of this work was carried out at the STFC ISIS Facility, UK, and part of this work was carried out at the Swiss Muon Source, Paul Scherrer Institut, Switzerland; we are grateful for the provision of beamtime.
The project was funded by EPSRC (UK) (Grant Nos: EP/N032128/1 and EP/N024028/1).
M. N. Wilson acknowledges the support of the Natural Sciences and Engineering Research Council of Canada (NSERC).
M. Gomil{\v s}ek would like to acknowledge Slovenian Research Agency under project Z1-1852.
We are grateful to B. Nicholson for help with sputtering the electrical contacts on the Cu$_2$OSeO$_3$ pellet, and F. Xiao for experimental assistance at PSI.
Research data from this paper will be made available via Durham Collections at \textcolor{red}{XXX}.

\bibliography{bib}

\end{document}